# Integrated Expert Recommendation Model for Online Communities

## Abeer El-korany [1]

Computer Science Department, Faculty of Computers & Information, Cairo University

## ABSTRACT

*Online communities have become vital places for Web 2.0 users to share knowledge and experiences. Recently, finding expertise user in community has become an important research issue. This paper proposes a novel cascaded model for expert recommendation using aggregated knowledge extracted from enormous contents and social network features. Vector space model is used to compute the relevance of published content with respect to a specific query while PageRank algorithm is applied to rank candidate experts. The experimental results show that the proposed model is an effective recommendation which can guarantee that the most candidate experts are both highly relevant to the specific queries and highly influential in corresponding areas.*

## INTRODUCTION

Knowledge sharing has been a research topic for the last decade. It was first mostly studied within organizational settings [9]. However, due to daily massive amount of knowledge and expertise sharing occurring online, knowledge sharing is being of considerable interest. Knowledge sharing environment includes repositories (containing those socially constructed as with Wikipedia [15]) as well as online forums designed for sharing knowledge and expertise. Discussion groups and forums, an emerging type of web-based communities for users to share knowledge and experiences or to provide social support, have attracted many users in various fields. The posted threads in discussion groups often contain users' professional and personal opinions, especially in some technical discussion groups. Each user can share experience and exchange knowledge by asking or answering questions. Through the website, members can post questions and then wait for answers, browse the questions that other users have asked, or search for answers to particular questions. Due to the increasing competition for users' discussion groups, different designs have emerged. Experts in discussion groups can earn a lot of prestige and sometimes economic interests by answering questions [18]. While, some sites allow anyone in the community to answer questions, others have individual "experts" filling that role; some charge askers and pay answerers. Others use leaderboards, points, or stars to encourage answering. For example, Microsoft gives awards to people who have made great contributions in the Office Discussion Groups every year. In Q&A sites it is possible that different types of questions and different rhetorical strategies will receive different responses depending on the community and its members. Furthermore, a newly joined user such communities, has no idea about how to ask an appropriate or to search these large questions/answers archives to retrieve high quality content. Therefore, expert identification in this special context is thus becoming a very important research problem. It is becoming necessary to automatically find experts in online community, in order to disseminate the newly posted questions to the appropriate experts, who

 

can provide high quality answers to these questions [5,19,23]. Accordingly, the overall answer quality is significantly improved.

Expert finding methods can be categorized according to the sources of knowledge and information into: domain knowledge driven methods and domain knowledge independent methods [34]. Domain knowledge driven methods involve domain knowledge to build user profiles, such as the content of documents authored by experts, and compute the relevance between the profiles and user input queries. On the other hand, reference information between experts is employed as domain independent knowledge to rank experts. This metadata created by explicit support for social interactions between users known as social features. These include comments, ratings using vote up, vote down and stars, as well as authorship and attribution information [12]. Such explicit feedback provides a strong indication of the content quality [1]. Existing research shows that both of the document contents and the reference relations are valuable information sources for expert finding.

In this paper, a novel expert recommendation model for online communities (EXREC) is proposed in which domain dependent and domain independent knowledge are combined. The former is mined from content published by the user and represent the user's domain knowledge level for a specific query. The latter is derived from network structure and social information and is used to determine the user' level of expertise. The model decomposes two cascaded phases. The first one, domain expert matching, through which information retrieval techniques are applied on discussion thread contents in order to determine relevance between user input queries (question) and user's historical question-answers. The second phase, the expert ranking phase utilizes online activities and user interactions to determine the level of expertise. It explores the network structure of the candidate experts (nodes) and define key node which present critical or important person. The novelty of our approach is that the discovery of experts is not only based on content (available as predefined user published posts) but also improved using knowledge extracted from historical interactions and exchanged knowledge in social context. To evaluate the proposed model, we choose the widely used stackoverflow which is a successful, active community of software developers that provide answers to each other's questions as our testing datasets. Through the analysis of this community we were able to define the features of a user that make them a qualified expert, and help new user with adequate set of candidate experts relevant to the posted query. The final experimental results show that EXREC outperforms existing methods are successfully filtered. The rest of this paper is organized as follows: Section 2 briefly reviews related literatures. Section 3 describes our expert identification approach in details. Section 4 depicts the evaluation methods and demonstrates the experimental results. Conclusion and future work are given in Section 5.

## RELATED WORKS

Expert finding is the task of finding users who can provide a large number of high quality, complete, and reliable answers [27]. These systems attempt to leverage the social network within an organization or community to help find the appropriate persons [35]. Many research works have been devoted to solve the expert finding problem in various contexts [1,31]. Existing approaches for finding experts in online community can be classified into link analysis and content analysis approaches. Recently, an attempt to integrate the two approaches is presented which is illustrated here.

### 1.1   Content-based approaches

The term expert is used to refer to a person/agent with a high degree of a skill or knowledge of a certain subject. Expert' supplied content is considered as main source to identify the level of expertise of an expert. Thus, the problem of expert identification can be solved using IR and NLP techniques such that relevance between queries and knowledge of experts could be identified. The work done in[3], presented two probability models to identify experts using available multilingual data from intranet of a research organization which cover a broad range of expertise areas. In [24] content of question/answer community could be classified using machine learning techniques and enhanced with semantic features which is extracted from a question.Other research work in [28] utilized document content, which provided high relevance search but cannot serve as direct evidence of expertise. Improved results were obtained by propagating the relevance following the candidate-document links. Other approaches applied topical analysis [13] using LDA model based on the latent topics embodied in document contents. This work has been applied on Yahoo! Answer. A few studies tried to employ other information sources to address the expert ranking problem such as employing of ontology. An expert finder system based on ontologies expressing skills of experts has been described in [6]. However, the work did not discuss how to obtain information regarding users' level of expertise in an automated manner.

## 1.2 Link-analysis approaches

In a research community, the network captures previous successful collaborations among scientists, Thus, graph Graph-based mining algorithms such as HITS [22] and PageRank [26] were studied in [14] to estimate the expertise of users. Furthermore, email exchange information has been used to identify interaction patterns between people [8,21]. Analysis of email-based interactions has been used to identify information flow metrics in the social graph. Hypertext Introduced Topic Selection (HITS) [22] method was also applied in [7] to build social network on email communication relations. The experimental results show improvements over content-based methods but are not convincing enough since they considered only a limited number of candidates.

Recently, studies have included both network and content in order to find candidate experts. For example, expert finding in an enterprise dataset [10] combine document contents and network information. Their work proves the feasibility of the combination of different types of information for finding appropriate experts. But their model needs a set of seed experts to build the community, which requires much domain knowledge. A topic-sensitive probabilistic model has been proposed in [36] which find the experts by taking into account both the link structure and the topical similarity among users. First, they defined the topics that users are interested in. Then PageRank algorithm was applied to measure the expert saliency score in order to find experts in real world data set from Yahoo! Answers.

The proposed model is different from all the above in that it provides a new cascaded methodology that utilize different information sources to perform query-based expert ranking followed by network scoring mechanism.

## AN INTEGRATED EXPERT RECOMMENDATION MODEL

Online communities have become an active research area. Due to millions of users and hundreds of millions of questions and answers, users may post similar or identical questions multiple times and the quality of answers varies extremely. Community for Question Answering [30], consists of three components: a mechanism for users to submit questions in natural language, a venue for users to submit answers to questions, and a community built around this exchange. Some of these sites are subject-specific such as Stack Overflow, which is used as our case study, limits its scope to questions about programming. Content-based expert identification approaches may reflect whether a person knows about a topic, but difficult to distinguish person's relative expertise levels. On the other hand, link analysis methods could reflect the likelihood of a user to provide a qualified answer to others. The proposed cascaded model for expert recommendation (EXREC) shown in figure1 aims to find appropriate experts to answer a given question in QA community. It starts by identifying experts based on relevance between the question and users' previous posts (both questions and answers). Next, candidate experts are filtered based on hidden information extracted from network pattern and interaction. Finally, online network features that cover helper' reputation and trust such as certifications between members are used to define the final list of top ranked expert for a given question. The significance of EXREC is that it uses both user's relevance to a question which represent a user's knowhow, user influence and reputation on the subjects related to the target question. This is accomplished by integrating the published contents and network features created from the question/answer relationships among users.

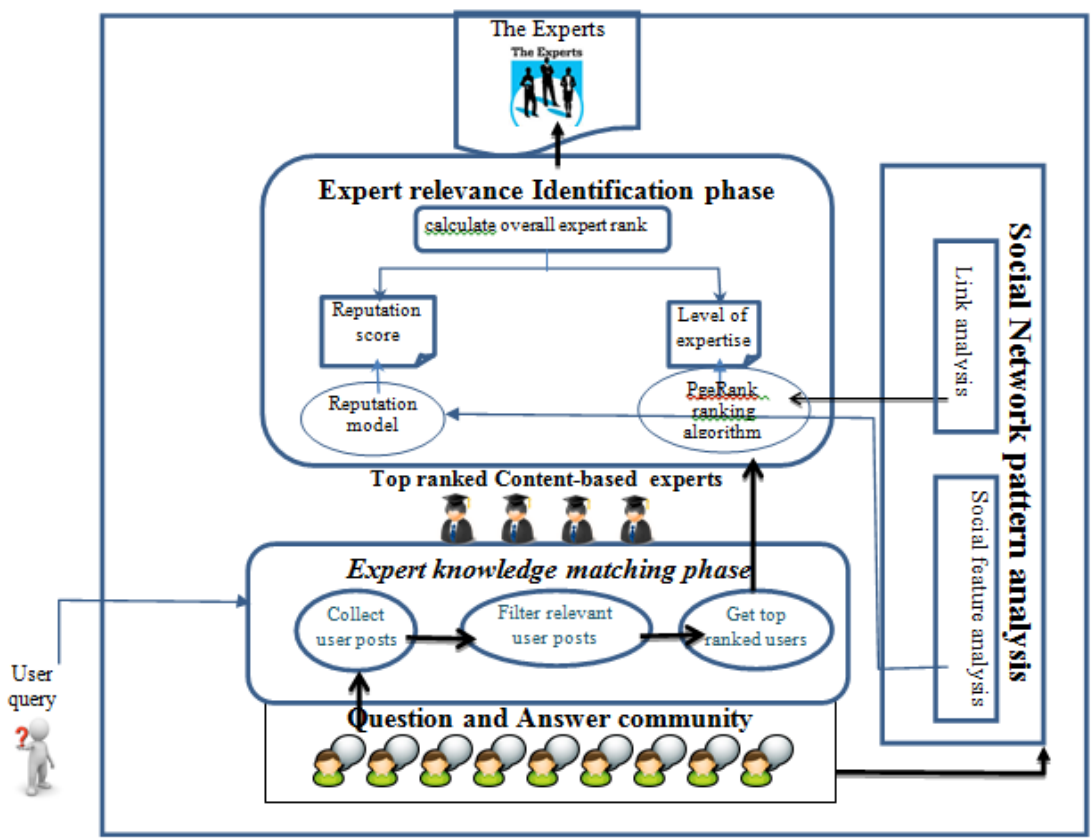

Figure1: Expert Recommendation Model

## 1.3 Expert knowledge matching

Online communities usually have a discussion thread structure [35]. A user posts a question, and then some other members post replies to either participate in the discussion or to answer a question posed in the original post. Approaches based on Information Retrieval and Natural Language Processing techniques [5,19,20] are suitable to be used here to identify experts who provide expertise in specific topic(s). These approaches could determine whom we want to send questions to, and whom can provide high quality information about specific topics. Thus, during expert matching phase expert(s) who has most related knowledge to a specific query are identified. For each member, all the previous posts she/he submitted is collected(in our case, all questions asked and answered provided by the user). A person's expertise is then described as a term vector and is used later for matching the submitted query with collected expertise using standard IR techniques. Vector space model is employed to calculate the similarity coefficient of the submitted query q and candidate experts using *TF-IDF* approach [29]. The expert relevance score is given as a commonly used cosine similarity between the query and aggregated expert knowledge related to that question. According to the algorithm shown in figure2, the result of expert matching phase is a list of ranked candidate relevance people. Expert similarity score identifies how similar a question to all community' users previous posts derived from term frequencies. Although this method measures the similarity between posted question and proposed answers, it cannot tell which expert has a stronger popularity or the highest influences in the network. Therefore, top ranked 20 candidate experts are used as a nominated list for the next phase.

```
Create initial-ranked expert lists (expert matching phase)
Users: List of all users
Posts: List of all posts,
APost: List of all answered posts,
BOWU: bag of words of each user
Tposts, Tusers, IBOW, Testset, Trainset, PSIM: empty list
Q: input query
U:= content vector of each user
For every post in the APost
Begin
            Calculate cosine  similarity between Q and APost
            PSIM:get  top 50 similar query
 End
 For each answer in PSIM
 Begin:
            Tusers:= get user ID of that answer
            TEstset= about 20% of BOWU of Tusers
            Trainset= about 80% of BOWU of Tusers
            identify similarity between Q and  Trainset
            USIM= get  top 20 similar users
  End
 For each user in USIM
 Begin:
            Calculate precision using TEstset
  End
            Rank USIM according to precision
```

Figure2: Expert matching Algorithm

## 1.4    Identify Expert relevance

The expert knowledge matching phase provides the initial list of candidate experts based on expertise extracted from their delivered answers and posts. However, users' reputation and influence inside the network should also be considered. Expert level of expertise is determined through structural and network information which will be illustrated in the following sub-sections. Structure features which indicate contributions of users to continuously generate rich online content and communication level with others in online community.  While other features such as prestige and trust are used for measuring and rankings key person.

### 1.4.1   Link-analysis model

Approaches used in expert identification based on link analysis, such as HITS and PageRank, utilize the network structure of web communities, to rank users. They rely on the fact that: direction of the links carries more information than just shared content. A user replying to another user's question usually indicates that the replier has superior expertise on the subject than the asker. Furthermore, weight of the links between users in online community reflects the likelihood of one user providing an answer to the others [25]. Accordingly, we utilize PageRank algorithm [26] to identify notable graph nodes (in our case list of candidate experts generated from **previous phase**) based on the graph features. The basic idea of applying PageRank algorithms is to utilize links features for each node (represents experts) such as frequency of communications based on their correspondence in the underlying domain. Characteristics of the relationships is the investigation of the attributes of the communication between two users (in particular its time and frequency) [32]. Based on the fact that experts answer a lot of questions and ask very few questions, features such as number of replies among users, user ratings, and common tagging are example of link features that could be used [32]. Using a modified version ExpertiseRanka algorithm [35], we generate a measure (expert level score) that considers not only the number of other users one helped, but also whom she/he helped by applying the following equation:

$$ER(A) = (1-d) + d\ (ER(U_1)/C(U_1) + \dots + ER(U_n)/C(U_n))$$

Where:
$C(U_i)$ : the total number of users answering $U_1$,
$D$ is a damping factor set d to 0.852
Users with high expert level score will tend to be the top candidate to answer question of specific topic. In our context, expertise-level score is calculated based on average number of provided answers ($a$),.

### 1.4.2    User reputation model

Expertise is closely related to structural prestige measures and rankings in social network studies [35]. Accordingly, there are opportunities to make use of such network structures and features to rank people's expertise in online communities. In such directed networks, people who receive many positive choices are considered to be prestigious, and prestige becomes significant especially if positive choices are not reciprocated [33]. Several research work [1,16] has shown that reputation of users are good indicators of the quality and reliability of the content. User reputation score could be calculated by means of features which represent the user's authority and influence in the community. Those features would help to define important experts whose participation is recognized by other in online community [4]. Statistical analysis approaches use social features [17] that reflect opinions of others who have similar behavior. These type of features may be answerer's acceptance ratio (the ratio of best answers to all the answers that the answerer answered previously), the score assigned to a question or answer by others, the reputation of the users. Higher scored participants are *key persons* having potential in the context of their *significance* in discussions [11]. Analysis of stackoverflow attributes, some features have been identified to determine user significance score such as: average views of answers, and average vote, average score and average favourite value which will be illustrated in details in next section

## EXPERIMENTS

Stack Overflow is an online platform where users can exchange knowledge related to programming and software engineering tasks. StackOverflow publishes their data under the creative commons license. We use the April 2009 data dump provided by Stack Overflow. The document collection contains all the questions and answers posted on the web site between between February 18, 2009 and June 7, 2009. This enabled us to rebuild the database from the published xml files and analyze the raw data from the community. StackOverflow provides its registered user the ability to ask new questions and answer existing questions, "vote" questions and answers up or down, based on the perceived value of the post, score the answers based on correctness and the degree of satisfaction, and mark post as favorite. Users of Stack Overflow can earn reputation points as a person is awarded 10 reputation points for receiving an "up" vote on any of their answers [2].

## 1.5 Data set and Experiment set up

For our purposes, we use an xml representation for users' posts, which contains the actual text content of the posts, as well as other community features. It contains post type (i.e., either question or answer), ID of the user who created each post, creation date, other users who comments on that post. For question, a pointer to all its answers, and score obtained of each answer. It has been found that answer posts comprise ∼72 % of the posts in the dataset, meaning that the majority of text content) is located in the answer posts [2]. Stackoverflow allows its users to give any solved question an evaluation (positive, neutral, or negative) regarding the degree of satisfaction of question-answer pair. Accordingly, it considers an answer which got +15 score as an accepted answer. Furthermore, it provides the following counts for each question: view count (number of user who viewed this question), and favourite count(the number of users who consider this question of interest). For each extracted post, we clean the textual content of the extracted posts in four steps. First, we discard any code snippets that are present in the posts, remove all HTML tags, remove common English-language stop words such as "a", "the" and "is", which do not help to create meaningful topics, and finally, stemming is applied to map words to their base form. Moreover, we also maintain other metadata for each user and accordingly each user's profile would contain the following information:

(1)Terms extracted from questions and associated answers
(2) Users' network structure features (adjacency matrix)
(3) Users' reputation features (e.g., average views, average score, average favourite count, and reputation)

## 1.6 Evaluation Metrics

To evaluate the cascaded model for expert recommendation, we use the widely studied metrics in information retrieval which are:

1. *Average Precision@n* (Avg. P@n): This metric denotes the average ratio of the relevant experts in top *n* identified experts for each query

2. *Mean Average Precision* (MAP): This metric is the mean of the average precision scores for each query.

We want to measure the accuracy of the expert identification system in two ways. First, we evaluate the accuracy of the expert matching phase (which considers the domain knowledge of the experts) by calculating the precision of retrieved answers. Thus, answers for each user are randomly divided into a training and a test set (80 and 20 % respectively). A set of queries have been applied in form of question with length varies from 10 to 20 term. Top ranked experts with respect to each question are retrieved. Precision is calculated using similarity score between terms in each query and bag of terms of each user. As shown in figure3, precision values have the same pattern for each of the provided questions. Furthermore, precision for the top ranked expert in five different questions was 1 which indicates the effectiveness and accuracy of the expert matching phase. In order to improve the diverge of ranked expert, we apply other set of experiments to measure the whole model

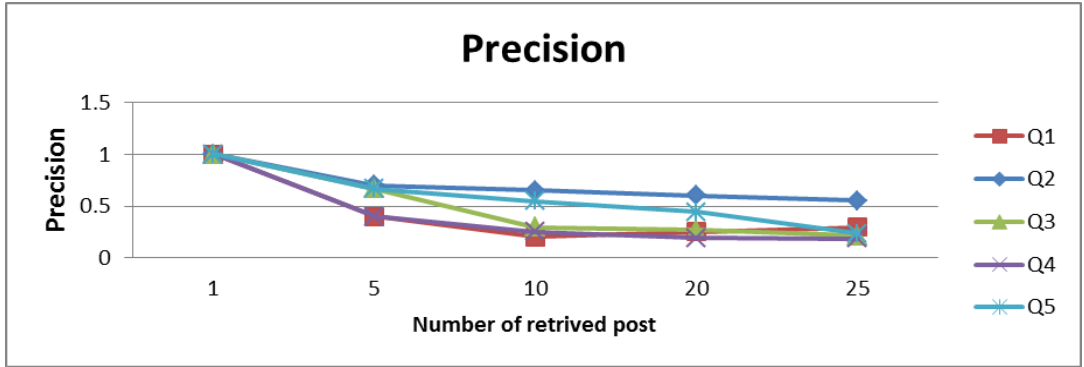

Figure3: Precision of retrieved users for different questions

Second, we compare the ranked experts obtained from phase2 with actual user acceptance ratio in the community. According to [2], answers that obtained score more than 15 are considered as accepted answers. Thus, to assess the quality of answers of each recommended expert, we compute the average score of the test set she/he has. By comparing the average score obtained from test set with *Mean Average Precision (*MAP *)*of the training set. MAP is calculated using equation1:

$$MAP = \frac{1}{|Q|}\left(\sum_{Q_i}\frac{1}{|R_i|}\left(\sum_{j=1}^{n}rel(D_j)\frac{\sum_{k=1}^{j}rel(D_k)}{r_j}\right)\right)$$  Equ1

This type of comparison provides an indication of the proportion of relevant retrieved answers, with respect to the actual score of all answers (level of expertise). Table2 presents this comparison between the MAP value obtained from applying five different questions to the whole system at the following level $P_1$, $P_5$, $P_{10}$, $P_{20}$. These values represent the accuracy of EXREC model for top ranked experts. The last column in table1 is the average accepted answers of those experts with respect to the whole answers which reflect others acceptance to her/his answers. Values shown in table1 guarantee the effectiveness of the proposed model as the value of MAP is directly proportional with the actual expertise of top ranked experts

| Query | MAP | Percentage of user acceptance answers |
|-------|--------|---------------------------------------|
| Q1 | 0.4288 | 0.21 |
| Q2 | 0.701 | 0.3 |
| Q3 | 0.488 | 0.22 |
| Q4 | 0.405 | 0.15 |
| Q5 | 0.5796 | 0.24 |

Table1: Comparison between MAP for top ranked expert and their actual acceptance in the QA community

In order to verify the effectiveness of the cascade model, we compare the expert ranked list from phase1 and the ranked list obtained from whole system after applying phase2 with the actual rank obtained from communities in form of current reputation score. According to table2, expertID1 has got the fourth rank from after applying phase1. This rank has been escalating after applying the whole model and thus expert1 got the first place among his relevant. This result matches with the actual reputation expert1 has from her/his colleagues in the community. This indicates that the cascaded model would converge the results to be matched the actual tangible value. Accuracy of the proposed model is also proven with the other top five ranked experts.

|  | expert-matching phase rank | Expert –relevance phase rank | | Actual reputation value |
|---|---|---|---|---|
| Expert1 | 4 | 1 | 180.196345 | 34638 |
| Expert 2 | 1 | 2 | 153.28205 | 4790 |
| Expert 3 | 2 | 3 | 109.870966 | 3489 |
| Expert4 | 3 | 4 | 47.248406 | 3216 |
| Expert5 | 5 | 5 | 10.254218 | 995 |

Table2: Comparison between rank obtained from each separate phase and real reputation of experts
.

## CONCLUSION AND FUTURE WORK

This research proposes a model for expert recommendation in online community (EXREC). Unlike other conventional approaches, expertise level is identified through two main folds. The first one is subject relevance which is expressed by the content obtained from by the users' discussion and facilitate the identification of most relevant experts to specific queries.

While the second fold utilizes information extracted based on link-based features that cover user's reputation and influence. Link-based features significantly improves the expert ranking and expert reputation is also used to converge the level of expertise. Possible improvements would be using domain ontology to better identify user's knowledge and extract topics of interest based on the ontology. Furthermore, team of experts in specific topic could be identified through building expert networks on the specific references contained in the contexts, like citations in publications.

## Author

Abeer El_korany received Ph.D. in Electronics and Communications Engineering, 2002, M.Sc. in Electronics and Communications Engineering 1996, B.S. in <u>EE</u> '92 from Cairo University, faculty of engineering Egypt. She was a key researcher in the expert system development from 93-2003 at center laboratory of agriculture expert system (CLEAS) before joining department of computer science faculty of computer and information , cairo university. Her research interests include quality assurance of knowledge-based system, ontology engineering, knowledge management, semantic web,social network, recommender systems

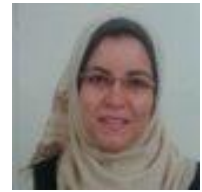